\documentclass[reprint,superscriptaddress,showpacs,amsmath, amssymb,aps,pra]{revtex4-1}
\usepackage{amssymb}
\usepackage{graphicx}
\usepackage{dcolumn}
\usepackage{dsfont}
\usepackage{bm}
\usepackage{subfigure}
\usepackage[ruled]{algorithm2e}
\usepackage{hyperref}
\usepackage{appendix}
\usepackage{ulem}
\usepackage{color}

\hypersetup{colorlinks=true, citecolor=blue, urlcolor=blue, linkcolor=blue}
\begin{document}
\title{Variational quantum algorithms for dimensionality reduction and classification}
\author{Jin-Min Liang}
\author{Shu-Qian Shen}
\email{sqshen@upc.edu.cn.}
\author{Ming Li}
\author{Lei Li\\
{\small{\it College of Science, China University of Petroleum, 266580 Qingdao, P.R. China}}\\
(Published 16 March 2020)}
\begin{abstract}
In this work, we present a quantum neighborhood preserving embedding and a quantum local discriminant embedding for dimensionality reduction and classification. We demonstrate that these two algorithms have an exponential speedup over their respectively classical counterparts. Along the way, we propose a variational quantum generalized eigenvalue solver that finds the generalized eigenvalues and eigenstates of a matrix pencil $(\mathcal{G},\mathcal{S})$. As a proof-of-principle, we implement our algorithm to solve $2^5\times2^5$ generalized eigenvalue problems. Finally, our results offer two optional outputs with quantum or classical form, which can be directly applied in another quantum or classical machine learning process.
\end{abstract}
\maketitle
\section{Introduction}
Dimensionality reduction is significant to many algorithms in pattern recognition and machine learning. It is intuitively regarded as a process of projecting a high-dimensional data to a lower-dimensional data, which preserves some information of interest in the data set \cite{Sarveniazi2014,Sorzano2014}. The technique of dimensionality reduction has been variously applied in a wide range of topics such as regression \cite{Hoffmann2009}, classification \cite{Vlachos2002}, and feature selection \cite{Chizi2010}.

Broadly speaking, all of these techniques were divided into two classes: linear and non-linear methods. Two most popular methods for linear dimensionality reduction are principal component analysis (PCA) and linear discriminant analysis (LDA). PCA is an orthogonal projection that minimizes the average projection cost defined as the mean squared distance between the data points and their projections \cite{PCA}. The purpose of LDA is to maximize the between-class variance and minimize within-class scatter when the data has associated with class labels \cite{LDA}. The most popular algorithm for non-linear dimensionality reduction is manifold learning \cite{Cayton2005}. The manifold learning algorithm aims to reconstruct an unknown nonlinear low-dimensional data manifold embedded in a high-dimensional space \cite{ML2012}. A number of algorithms have been proposed for manifold learning, including Laplacian eigenmap \cite{LE2001}, locally linear embedding (LLE) \cite{LLE2000}, and isomap \cite{Iso2000}. Manifold learning has been successfully applied for video-to-video face recognition \cite{Hadid2009}. These nonlinear methods consider the structure of the manifold on which the data may possibly reside compared with kernel-based techniques (e.g., kernel PCA and kernel LDA).

We are witnessing the development of quantum computation and quantum hardware. The discovery of quantum algorithm for factoring \cite{Shor1994}, database searching \cite{Grover1996} and matrix inverse \cite{HHL2009} has shown that quantum algorithms have the capability of outperforming existed classical counterparts. Recently, quantum information combines ideas from artificial intelligence and deep learning to form a new field: quantum machine learning (QML) \cite{QML2017}. For classification and regression, QML algorithms \cite{QSVM2014,Wiebe2015,LR2016,LR2017,Aimeur2013} also have shown advantages over their classical machine learning algorithms. However, much algorithms rely on the large-scale, fault-tolerate, universal quantum computer which may be achieved in the distant future. Specifically, these algorithms will require enormous number of qubits and long depth of circuit to achieve quantum supremacy.

Fortunately, noisy intermediate-scale quantum (NISQ) devices are thought of as a significant step toward more powerful quantum computer \cite{NISQ2018}. This NISQ technology will be available in the near future. In this setting, hybrid algorithmic approaches demonstrate quantum supremacy in the NISQ era. This hybridization reduces the quantum resources including qubit counts, numbers of gates, circuit depth, and numbers of measurements \cite{Larose2019}. Variational hybrid quantum-classical algorithms aim to tackle complex problems using classical computer and near term quantum computer. The classical computer finds the optimal parameters by minimizing the expectation value of objective function which is calculated entirely on the quantum computer.

The first class variational quantum algorithms have been proposed for preparing the ground state of a Hamiltonian \cite{VQE2014}. For a Hamiltonian $\mathcal{H}$ which is too large to diagonalize, one can approximate the ground state of the given Hamiltonian using the Rayleigh-Ritz variational method. After parametrizing the trial quantum states, one can perform an optimization subroutine to find the optimal state by tuning the optimal parameter. Variational method is also applied to obtain the excited state of a Hamiltonian \cite{Higgott2019,Jones2019} and diagonalize a quantum state \cite{Larose2019}.
Another class of hybrid algorithms is designed to find application in machine learning including the quantum approximate optimization algorithm (QAOA) \cite{QAOA2014}, variational quantum algorithms for nonlinear partial differential equations \cite{Lubasch2019}, and linear systems of equations \cite{Xu2019,An2019,Huang2019,Carlos2019}.

Inspired by the significant advantage of quantum algorithms, some authors designed quantum algorithms to reduce the dimension of a large data set in high-dimensional space. Quantum principal component analysis (qPCA) \cite{QPCA2014} and quantum linear discriminant analysis (qLDA) \cite{QLDA2016} are two potential candidates capable of compressing high-dimensional data set and reducing the runtime to be logarithmic in the number of input vectors and their dimensions. These two protocols yield global mappings for linear dimensionality reduction and obtain the projected vectors with only quantum form. Thus a complicated quantum tomography \cite{Nielsen2000} is needed if one would like to know all information of the projected vectors.

Motivated by manifold learning and quantum computation, one natural question arises of whether there have a quantum algorithm for dimensionality reduction and pattern classification, and in which preserves the local structure of original data space. To tackle this issue, we present two variational quantum algorithms. First one is quantum neighborhood preserving embedding (qNPE) which defines a map both on the training set and test set. The core of qNPE is a variational quantum generalized eigensolver (VQGE) based on Rayleigh quotient, a variant of quantum variational eigenvalue solver (QVE) \cite{VQE2014}, to prepare the generalized eigenpair $(\lambda, x)$ of the generaliezd eigenvalue problem $Ax=\lambda Bx$. Based on the presented VQGE, we propose a quantum version of local discriminant embedding \cite{LDE2005} for pattern classification on high-dimensional data. We show that these two algorithms achieve an exponential speedup over their classical counterparts.

The organization of the paper is as follows. In Sec. II, we give a quantum neighborhood preserving embedding (qNPE) for dimensionality reduction. The numerical experiments are conducted using five qubits to demonstrate the correctness of VQGE in subsection E of Sec. II. In Sec. III, we introduce the quantum local discriminant embedding (qLDE) in detail for classification problem. A summary and discussion are included in Sec. IV.
\section{Quantum Neighborhood Preserving Embedding}
Local linear embedding (LLE) \cite{LLE2000} is an unsupervised method for nonlinear dimensionality reduction; thus it does not evaluate the maps on novel testing data points \cite{NPE2005}. Neighborhood preserving embedding (NPE) is thought of as a linear approximation to the LLE algorithm \cite{NPE2005}. NPE tries to find a projection suitable for the training set and testing set. Different from other linear dimensional reduction methods (PCA and LDA), which aim at maintaining the global Euclidean structure, NPE preserves the local manifold structure of data space. We assume that the regions will appear to be locally linear when the size of neighborhood is small and the manifold is sufficiently smooth. Experiments on face recognition have been conducted to demonstrate the effectiveness of NPE \cite{NPE2005}. Here, we introduce a quantum neighborhood preserving embedding (qNPE). Given a set of points $\{x_i\}_{i=0}^{M-1}\in\mathcal{M}$ and $\mathcal{M}$ as a nonlinear manifold embedded in a $D$-dimensional real space $\mathcal{R}^{D}$, our qNPE attempts to retain the neighborhood structure of the manifold by representing $x_i$ as a convex combination of its nearest neighbors. In particular, qNPE finds a transformation matrix $A$ that maps these $M$ points and test point $x_{test}$ into a set of points $y_0,y_1,\cdots,y_{M-1},y_{test}\in\mathcal{R}^d$ in a lower-dimensional manifold space, where $y_i=A^{\dag}x_i$, $y_{test}=A^{\dag}x_{test}$, $d\ll D$, and the superscript $\dag$ denotes the conjugate transpose.

In the quantum setting, a quantum state preparation routine is necessary to construct the quantum states $\{|x_i\rangle\}_{i=0}^{M-1}$ corresponding to classical vectors $\{x_i\}_{i=0}^{M-1}$. Assume that we are given oracles for data set $\{x_i|x_i\in\mathcal{R}^D\}_{i=0}^{M-1}$ that return quantum states $\{|x_i\rangle\}_{i=0}^{M-1}$. Mathematically, an arbitrary $D-$dimensional vector $\vec{x}_i=(x_{i0},x_{i1},\cdots,x_{i(D-1)})^{\dag}$ is encoded into the $D$ amplitudes $\{x_{i0},x_{i1},\cdots,x_{i(D-1)}\}$ of an $O(\log_2 D)$-qubits quantum system, $|x_i\rangle=\sum_{j=0}^{D-1}x_{ij}|j\rangle$, where $\{|j\rangle\}$ is the computational basis \cite{LR2016}.
\subsection{Find the $K$-nearest neighbors}
The first step of qNPE is the construction of a neighborhood graph according to the given data set. The construction of an adjacency graph $G$ with $M$ nodes relies on the $K$ nearest neighbors of $x_i$. If $x_j$ is one of the $K$ nearest neighbors of $x_i$, then a directed edge will be drawn from the $i$th node to the $j$th node; otherwise, there is no edge. To preserve the local structure of the data set, we first develop an algorithm (Algorithm 1) to search the $K$ nearest neighbors of point $x_i$.

Some notations are needed to understand Algorithm 1. Let $\{f(i)|i\in[0,1,\cdots,M-1]\}$ be an unsorted table of $M$ items. We would like to find $K$ indexes set $\mathcal{N}=\{j_{1},j_{2},\cdots,j_K\}$ of the element such that $f(j_1)\leq f(j_2)\leq\cdots\leq f(j_K)\leq f(j)$ where $\{j_{1},j_{2},\cdots,j_K,j\in[0,1,\cdots,M-1]\}$ and $j\notin \mathcal{N}$. We call it quantum $K$ nearest neighbors search which is a direct generalization of the quantum algorithm for finding the minimum \cite{Durr1996}. One of our results is the following theorem.

\textit{Theorem 1.} For a given quantum state set $\{|x_i\rangle\}_{i=0}^{M-1}$,
let $[0,1,\cdots,M-1]$ be an unsorted database of $M$ items, each holding an inner product value. Algorithm 1 finds all lower $K$ indexes with probability at least $\frac{1}{2}$ costing
$$O\Bigg(\frac{M(M-1)}{2}\log_2 D\Bigg),$$
with query complexity $O(KM\sqrt{M}).$

\textit{Proof.} Quantum $K$ nearest neighbors search tries to find the $K$ lower values of an unsorted data set. In step 1, given a state set $$\left\{|x_i\rangle=\sum_{j=0}^{D-1}x_{ij}|j\rangle\right\}_{i=0}^{M-1},$$
we first estimate the square of inner product $|\langle x_i|x_k\rangle|^2$ over all data points for $i,k=0,1,\cdots,M-1$ via swap test each running costs $O(\log_2 D)$ \cite{swaptest}.
The number of performing swap test is
$$T_{swap}=\sum_{i=0}^{M-1}i=\frac{M(M-1)}{2}.$$
Thus the overall cost of estimating square of inner product is $O(\frac{M(M-1)}{2}\log_2 D)$.

In steps 2-4, we find $K$ lower index set $\mathcal{N}$ of one point $|x_i\rangle$. By adjusting $s=s-1$, the index set $T_s$ deletes one element every times. We repeat $K$ times on the updated index set $T_s$ to obtain the $K$ lower index $\mathcal{N}$ mapping to $K$ smallest values. D\"{u}rr and H\o yer \cite{Durr1996} have shown the query complexity of finding the minimum value is $O(\sqrt{M})$. In our algorithm, the query complexity of finding $K$ nearest neighbors of one state $|x_i\rangle$ is
\begin{equation}
O\Bigg(\sum_{k=1}^{K}\sqrt{M-(k-1)}\Bigg)<O\Bigg(K\sqrt{M}\Bigg),
\end{equation}
which has an upper bound $O(K\sqrt{M})$. Thus, the overall query complexity of traversal all $M$ quantum state have an upper bound $Q=O(KM\sqrt{M})$.$\hfill\blacksquare$

Contrasting this to the situation where the entire algorithm is applied on a classical computer, we require exponential resources in both storage and computation. Firstly, storage of the quantum state $|x\rangle$ using the known quantum encoding technique requires $2^D$ complex numbers. Moreover, each distance calculation is $O(D)$, and thus the time complexity is $O(\frac{MD(M-1)}{2})$. Finally, the classical search complexity is $O(M)$ for an unsort data set containing $M$ elements. For our $K$ nearest-neighbor algorithm, the overall search complexity has an upper bound $O(KM^2)$. The computational complexity of the classical nearest-neighbor algorithm has been analysed in \cite{duba2012pattern}. We roughly estimate that the time complexity is $O(\frac{MD(M-1)}{2})$ and the query complexity is $O(KM^2)$. It is clear to see that the quantum $K$ nearest neighbours search achieves an exponential speedup in the dimensionality of quantum states.
\begin{algorithm}[h]
\caption{Quantum $K$ nearest neighbors search}
\LinesNumbered
\textit{step 1:} Estimate the overall square of inner product value via swap test with error at most $\epsilon$ costing $O(\frac{M(M-1)}{2}\log_2 D)$.\\
\textbf{Repeat the following steps $K$ times:}\\
\textit{step 2:} Define an index set $T_s=[0,1,\cdots,s-1]$ where $s$ is initialized as $M$.\\
\textit{step 3:} Apply the minimum searching algorithm \cite{Durr1996} and output a minimum index $j$ in runtime $O(\sqrt{s})$ with probability at least $\frac{1}{2}$.\\
\textit{step 4:} Delete the minimum index and reset $s=s-1$.\\
\textbf{Outputs:} $\mathcal{N}=\{j_1,j_2,\cdots,j_K\}$.
\end{algorithm}

The query complexity of the presented Algorithm 1 can be further reduced to $O(M\sqrt{KM})$ using the idea of \cite{Durr2006,Miyamoto2019}. D\"{u}rr \textit{et al.} \cite{Durr2006} transformed the problem of finding $d$ smallest values to find the position of the $d$ zeros in the matrix consisting of Boolean matrices with a single zero in every row, which can be seen as a part of graph algorithm. Different from \cite{Durr2006}, Miyamoto and  Iwamura \cite{Miyamoto2019} first found a good threshold by quantum counting and then values of all $d$ indices are found via amplitude amplification. The values of all $d$ indices are less than the value of the threshold index.

In summary, Algorithm 1 finds the $K$ nearest neighbors $\mathcal{N}_i=\{x_0^i,x_1^i,\cdots,x_{K-1}^i\}$ of quantum state $|x_i\rangle$ \cite{nearest}. The presented algorithm is based on two algorithms: finding minimum and swap test. First, we reformulate the algorithm for finding $K$ indices by updating the search set. Secondly, we explicitly analyse the time complexity and query complexity.

For implementation of the quantum $K$ nearest neighbours search, only one free parameter, $K$, is taken into account. The threshold $K$ affects the performance of qNPE. Specifically, it remains unclear how to select the parameter $K$ in a principled manner. The qNPE will lose its nonlinear character and behave like traditional PCA if $K$ is too large. In this case, the entire data space is seen as a local neighbourhood. Moreover, if the threshold $K$ is bigger than the dimension of data point, the loss function (2) described in subsection B will have infinite solutions and the optimal question will be irregular.
\subsection{Obtain the weight matrix}
Let $W$ denote the weight matrix with element $\omega_j^i$ having the weight of the edge from node $i$ to node $j$, and 0 if there is no such edge. For maintaining the local structure of the adjacency graph, we assume each data node can be approximated by the linear combination of its local neighbor nodes \cite{NPE2005}. It is the weight matrix that characters the relationship between the data points. The weights can be calculated by the following convex optimization problem,
\begin{equation}
\begin{aligned}
&\min\quad \Phi(\omega_j^i)=\sum_{i=0}^{M-1}\left\|x_i-\sum_{j=0}^{K-1}\omega_j^ix_j^i\right\|^2\\
&\textrm{such}\quad\textrm{that}\quad \sum_{j=0}^{K-1}\omega_j^i=1,i=0,1,\cdots,M-1.
\end{aligned}
\end{equation}
Using the Lagrange multiplier to enforce the constraint condition $\sum_{j}\omega_j^i=1$, the optimal weights are given by:
\begin{equation}
\begin{aligned}
\omega_i&=(\omega_0^i,\omega_1^i,\cdots,\omega_{K-1}^i)^{\dag}=\frac{G_i^{-1}\vec{1}}{\vec{1}^\dag G_i^{-1}\vec{1}},
\end{aligned}
\end{equation}
where the covariance matrix is defined as $G_i=\mathcal{A}_i^{\dag}\mathcal{A}_i$, and $\mathcal{A}_i=X_i-N_i\in\mathcal{R}^{D\times K}$, $X_i=(x_i,x_i,\cdots,x_i)\in\mathcal{R}^{D\times K}$, $\vec{1}=(1,1,\cdots,1)^\dag\in\mathcal{R}^{K}$, and $N_i=(x_0^i,x_1^i,\cdots,x_{K-1}^i)\in\mathcal{R}^{D\times K}$. The column vector $x_j^i\in\mathcal{N}_i$ of $N_i$ represents the $K$-nearest data points close to the data point $x_i$. The detailed derivation of Eq. (3) is shown in \cite{LLE2000}. Each $K$-nearest data point is in a $D$-dimensional real space $\mathcal{R}^D$. Our goal is to find weight quantum state $|\omega_i\rangle$ that satisfies
\begin{equation}
\begin{aligned}
|\omega_i\rangle\propto|G_i^{-1}\vec{1}\rangle=\frac{G_i^{-1}\vec{1}}{\|G_i^{-1}\vec{1}\|}.
\end{aligned}
\end{equation}
A key idea is to find the inverse of the matrix $G_i$ with quantum technique. If the weight is not unique, some further regularization should be imposed on the cost function of Eq. (2) \cite{LLE2000}.

In the following process, we make use of the matrix inverse algorithm shown in \cite{HHL2009,QSVD2018} to prepare the quantum state $|\omega_i\rangle$. Let the singular value decomposition (SVD) of $\mathcal{A}_i$ be $\mathcal{A}_i=U\Sigma V^{\dag}=\sum_j\sigma_j^i|u_j^i\rangle\langle v_j^i|$, then the eigenvalue decomposition of covariance matrix $G_i$ \cite{Explanation1} is
\begin{equation}
\begin{aligned}
G_i=\sum_{j=0}^{K-1}(\sigma_j^i)^2|v_j^i\rangle\langle v_j^i|.
\end{aligned}
\end{equation}
Thus $|G_i^{-1}\vec{1}\rangle$ can be reexpressed as
\begin{equation}
\begin{aligned}
|G_i^{-1}\vec{1}\rangle=\sqrt{\frac{1}{\sum_{j=0}^{K-1}|\beta_j^i|^2/|\sigma_j^i|^4}}\sum_{j=0}^{K-1}\frac{\beta_j^i}{(\sigma_j^i)^2}|v_j^i\rangle,
\end{aligned}
\end{equation}
where $\beta_j^i=\langle v_j^i|\vec{1}\rangle$. Assume that we are given a matrix oracle $O_i$ which accesses the element $\mathcal{A}_{mn}^i$ of the matrix $\mathcal{A}_i$:
\begin{equation}
|m\rangle |n\rangle|0\cdots0\rangle\mapsto|m\rangle |n\rangle|\mathcal{A}_{mn}^i\rangle=|m\rangle |n\rangle|x_m^i-x_{mn}^i\rangle.
\end{equation}
This oracle $O_i$ can be provided by quantum random access memory (qRAM) using $O(KD)$ storage space in $O(\log_2^2 \max(K,D))$ operations \cite{qRAM2008}. With these preparations, we are able to efficiently simulate the unitary $e^{\imath\hat{\mathcal{A}}_i}$ and prepare the weights state $|\omega_i\rangle$, where $$\hat{\mathcal{A}}_i=\begin{pmatrix}0 & \mathcal{A}_i\\ \mathcal{A}_i^{\dag} & 0\end{pmatrix}.$$

To understand our algorithm quickly, we will give some details below. First of all, we perform quantum singular value decomposition (QSVD) of the matrix $\mathcal{A}_i$ on an initial state $|0\cdots0\rangle|\vec{1}\rangle$ to obtain the state $\sum_j\beta_j^i|\sigma_j^i\rangle|v_j^i\rangle$ containing singular values and right singular vectors of $\mathcal{A}_i$. The first register is assigned to store the singular values and the second register to decompose $|\vec{1}\rangle$ in the space spanned by the right singular vectors of $\mathcal{A}_i$.
The quantum state $|\vec{1}\rangle=\sum_{j=0}^{K-1}\frac{1}{\sqrt{K}}|j\rangle$ can be easily prepared by applying $O(\log K)$ Hadamard gates on $O(\log K)$ qubits $|0^{\otimes\log K}\rangle$. Mathematically,
\begin{equation}
\begin{aligned}
H^{\otimes\log_2K}|0^{\otimes\log_2K}\rangle&=\frac{1}{(\sqrt{2})^{\log_2 K}}(|0\rangle+|1\rangle)^{\otimes\log_2 K}\\
&=\sum_{j=0}^{K-1}\frac{1}{\sqrt{K}}|j\rangle.
\end{aligned}
\end{equation}

Now, we apply a unitary transformation taking $\sigma_j^i$ to $\frac{C_i}{|\sigma_j^i|^2}\sigma_j^i$, where $C_i$ is a normalized constant. Actually, this rotation can be realized by applying $R_y(2\arcsin\frac{C_i}{|\sigma_j^i|^2})$ \cite{Cao2013quantum,Duan2018efficient} on the ancilla qubit $|0\rangle$,
\begin{equation}
\begin{aligned}
&\sum_{j=0}^{K-1}\beta_j^i|\sigma_j^i\rangle|v_j^i\rangle|0\rangle\\
&\stackrel{R_y}{\longrightarrow}
\sum_{j=0}^{K-1}\beta_j^i|\sigma_j^i\rangle|v_j^i\rangle\Bigg(\frac{C_i}{|\sigma_j^i|^2}|1\rangle+\sqrt{1-\frac{C_i^2}{|\sigma_j^i|^4}}|0\rangle\Bigg).
\end{aligned}
\end{equation}
Next, uncompute the singular value register and measure the ancilla qubit to obtain 1. The system are left with a state proportional to
\begin{equation}
\begin{aligned}
|\omega_i\rangle&\propto\sqrt{\frac{1}{\sum_{j=0}^{K-1}|C_i\beta_j^i|^2/|\sigma_j^i|^4}}\sum_{j=0}^{K-1}\frac{C_i\beta_j^i}{|\sigma_j^i|^2}|v_j^i\rangle.
\end{aligned}
\end{equation}

It is clear to see that the weight states $\{|\omega_i\rangle\}_{i=0}^{M-1}$ can be prepared by repeating the above process $M$ times separately with the gate resources scaling as $O(MT_g)$, where $T_g$ denotes the number of required gate in the process of preparing the state $|\omega_i\rangle$. However, taking into account the extraction of embedding vectors requiring a reconstructed weight matrix, we introduce an improved approach which achieves a parallel speedup in the preparation of the weight matrix. We reconstruct the weight matrix $W=(|\omega_0\rangle,|\omega_1\rangle,\cdots,|\omega_{M-1}\rangle)$ via preparing a entanglement state $|\psi_W\rangle=\sum_{i=0}^{M-1}|\omega_i\rangle|i\rangle$. Theorem 2 validates the gate resources can be further reduced.

\textit{Theorem 2.} For a given quantum state set $\{|x_i\rangle\}_{i=0}^{M-1}$, the task of preparing $|\psi_W\rangle=\sum_{i=0}^{M-1}|\omega_i\rangle|i\rangle$ with error at most $\epsilon$ has runtime $$T_W=O\Bigg(\frac{\log^2(K+D)}{\epsilon^3}\sum_{i=0}^{M-1}\|\mathcal{A}_i\|_{max}^2\Bigg).$$
The required gate resources are $O(T_g+\log M)$.

\textit{Proof.} We add an ancilla $M$ dimension system which determines the applied unitary operator, given the initial state
$|\vec{1}\rangle_1|0\cdots0\rangle_2|0^{\otimes\log M}\rangle_3|0\rangle_4$. The register 3 gives the number of data set. After performing $O(\log_2 M)$ Hadamard gates on register 3, we apply the unitary operator
$$\Bigg(\sum_{i=0}^{M-1}U_i\otimes|i\rangle\langle i|\Bigg)\otimes\mathds{1}$$
on state
$$\sum_{i=0}^{M-1}|\vec{1}\rangle_1|0\cdots0\rangle_2|i\rangle_3|0\rangle_4,$$
where $U_i$ is the quantum phase estimation part of matrix $\hat{\mathcal{A}}_i$ and $\mathds{1}$ denotes the identity operator. This step obtains the state
\begin{equation}
\begin{aligned}
\sum_{i=0}^{M-1}\sum_{j=0}^{K-1}\beta_j^i|v_j^i\rangle_1|\sigma_j^i\rangle_2|i\rangle_3|0\rangle_4.
\end{aligned}
\end{equation}
And then rotate the singular value by applying $R_y(2\arcsin\frac{C_i}{|\sigma_j^i|^2})$ on the ancilla qubit $|0\rangle_4$. The system state is
\begin{equation}
\begin{aligned}
\sum_{i=0}^{M-1}\sum_{j=0}^{K-1}\beta_j^i|v_j^i\rangle_1|\sigma_j^i\rangle_2|i\rangle_3
\Bigg(\sqrt{1-\frac{|C_i|^2}{|\sigma_j^i|^4}}|0\rangle_4+\frac{C_i}{|\sigma_j^i|^2}|1\rangle_4\Bigg).
\end{aligned}
\end{equation}
Finally, uncomputing the second register and measuring the fourth register to see 1, we obtain the state
\begin{equation}
\begin{aligned}
\sum_{i=0}^{M-1}\sqrt{\frac{1}{\sum_{j=0}^{K-1}|C_i\beta_j^i|^2/|\sigma_j^i|^4}}
\sum_{j=0}^{K-1}\frac{C_i\beta_j^i}{|\sigma_j^i|^2}|v_j^i\rangle|i\rangle
\end{aligned}
\end{equation}
which is proportional to the entangled state $\sum_{i=0}^{M-1}|\omega_i\rangle|i\rangle$.

The runtime of preparing the state $\sum_{i=0}^{M-1}|\omega_i\rangle|i\rangle$ is dominated by the quantum singular value estimation of $\mathcal{A}_i\in\mathcal{R}^{D\times K}$. In the process, we consider an extended matrix $\hat{\mathcal{A}}_i\in\mathcal{R}^{(K+D)\times(K+D)}$ and obtain the eigenvalues of $\hat{\mathcal{A}}_i$ by performing quantum phase estimate. According to \cite{QSVD2018}, we prepare the state $|\omega_i\rangle$ with accuracy $\epsilon$ in runtime $O(\|\mathcal{A}_i\|_{max}^2\log_2^2(K+D)/\epsilon^3)$ where $\|\mathcal{A}_i\|_{max}$ is the maximal absolute value of the matrix elements of $A_i$. Therefore, the entangled state $\sum_{i=0}^{M-1}|\omega_i\rangle|i\rangle$ is prepared in runtime
\begin{equation}
T_W=O\Bigg(\frac{\log_2^2(K+D)}{\epsilon^3}\sum_{i=0}^{M-1}\|\mathcal{A}_i\|_{max}^2\Bigg).
\end{equation}

Overall, only extra $O(\log_2M)$ Hadamard gates are required along the way. Thus the quantum parallelism enables the gate resources to be reduced to $O(Tg+\log_2M)$ rather than $O(MTg)$.$\hfill\blacksquare$
\subsection{Variational quantum generalized eigenvalue solver}
In this subsection, we compute the linear projections $A$. The embedding of $x_i$ is accomplished by $y_i=A^{\dag}x_i$. Unlike PCA and LDA, we obtain the projection matrix $A$ by solving the following cost function based on the locally linear reconstruction errors:
\begin{equation}
\begin{aligned}
\Phi(y)=\sum_{i=0}^{M-1}\Bigg(y_i-\sum_{j=0}^{K-1}\omega_j^iy_j\Bigg)^2.
\end{aligned}
\end{equation}
Here, the fixed weights $\omega_j^i$ characterize intrinsic geometric properties of each neighborhood. Each high-dimensional data $x_i\in\mathcal{R}^D$ is mapped to a low-dimensional data $y_i\in\mathcal{R}^d,d\ll D$. The embedding vector $y_i$ is found by minimizing the cost function (15) over $y_i$. Following some matrix computation \cite{LLE2000,LDE2005}, the cost function can be reduced to the generalized eigenvalue problem:
\begin{equation}
\begin{aligned}
XQX^{\dag}a=\lambda XX^{\dag}a,
\end{aligned}
\end{equation}
where $X=(x_0,x_1,\cdots,x_{M-1}),Q=(I-W)^{\dag}(I-W)$, and $I=\textrm{diag}(1,\cdots,1)$. The detailed derivation is shown in \cite{NPE2005}.

The generalized eigenvalue problem, $\mathcal{G}x=\lambda \mathcal{S}x$, is an important challenge in scientific and engineering applications. Although Cong and Duan \cite{QLDA2016} has presented a Hermitian chain product to solve the generalized eigenvalue problem by replacing $\mathcal{S}^{-1}$ with $\mathcal{S}^{-1/2}$, the computation of matrix inverse is extremely difficult on classical computer.
Due to the above circumstances, Theorem 3 gives a variational quantum generalized eigenvalue solver (VQGE) for solving the generalized eigenvalue problem. Like the variational quantum eigenvalue solver (VQE) \cite{VQE2014}, our VQGE can also be run on near-term noisy devices. Algorithm 2 shows the outline of variational quantum generalized eigenvalue solver.

We first briefly review the subroutine quantum expectation estimation (QEE) \cite{VQE2014} in step 2 of Algorithm 2. The QEE algorithm calculates the expectation value of a given Hamiltonian $\mathcal{H}$ for a quantum state $|\varphi\rangle$. Any Hamiltonian can be rewritten as $M$ terms \cite{Berry2007,Childs2011,VQE2014}, for real parameter $h_{12\cdots}^{ij\cdots}$
\begin{equation}
\begin{aligned}\label{equ:Hamiltonian}
\mathcal{H}&=\mathcal{H}^1+\mathcal{H}^2+\cdots\\
&=\sum_{i1}h_1^i\sigma_1^i+\sum_{ij12}h_{12}^{ij}\sigma_1^i\otimes\sigma_2^j+\cdots,
\end{aligned}
\end{equation}
where indices $i,j,\cdots$ denote the subsystem on which the operator acts, and $1,2$ identify the Pauli operator. Each subitem $\mathcal{H}^{m}$ is a summation of some tensor products of Pauli operators. According to Eq. (17), the expectation value is
\begin{equation}
\begin{aligned}
\langle\mathcal{H}\rangle&=\langle\mathcal{H}^1\rangle+\langle\mathcal{H}^2\rangle+\cdots\\
&=\sum_{i1}h_1^i\langle\sigma_1^i\rangle+\sum_{ij12}h_{12}^{ij}\langle\sigma_1^i\otimes\sigma_2^j\rangle+\cdots.
\end{aligned}
\end{equation}
As a result, each expectation $\langle\mathcal{H}^m\rangle$ is directly estimated using fermionic simulations \cite{Ortiz2001} or statistical sampling \cite{Romero2019}.

\begin{algorithm}[]
\caption{Variational quantum generalized eigenvalue solver}
\LinesNumbered
\textit{step 1:} Design a quantum circuit $U(\vec\theta)$, controlled by a set of experimental parameters $\vec\theta=(\theta_i)$, which can prepare states $|\varphi\rangle=|\varphi(\{\theta_i\})\rangle$.\\
\textit{step 2:} Define a objective function $f(\{\theta_i\})=\frac{\langle\varphi|\mathcal{H}_\mathcal{G}|\varphi\rangle}{\langle\varphi|\mathcal{H}_\mathcal{S}|\varphi\rangle}$ which $f$ maps parameters to a Rayleigh quotient of $|\varphi\rangle$ if $\langle\varphi|\mathcal{H}_\mathcal{S}|\varphi\rangle\neq0$.\\
\textit{step 3:} Find all the generalized eigenvalues and corresponding generalized eigenstates.\\
\quad(a) Compute the expectation $\langle\mathcal{H}_{\mathcal{G}(\mathcal{S})}^1\rangle,\langle\mathcal{H}_{\mathcal{G}(\mathcal{S})}^2\rangle,\cdots$, on $|\varphi_n\rangle=|\varphi_n(\{\theta_i\})\rangle$ for all terms of $\mathcal{H}_{\mathcal{G}(\mathcal{S})}$ by quantum expectation estimation \cite{VQE2014}, which $n$ denotes the iteration times of repeating Step 3.\\
\quad(b) Sum these values with weights to obtain $$f_n=\frac{\langle\varphi_n|\mathcal{H}_\mathcal{G}|\varphi_n\rangle}{\langle\varphi_n|\mathcal{H}_\mathcal{S}|\varphi_n\rangle}.$$\\
\quad(c) Apply the classical minimization algorithm (e.g. gradient descent) to minimize $f_n$ and determine the new parameter $\{\theta_i^{n}\}$.\\
\quad(d) Using step 1 to generate the state $|\varphi_{n}\rangle=|\varphi_{n}(\theta_i^{n})\rangle$.\\
\textit{step 4:} Update the Hamiltonian:\\
\quad(a) if $\mathcal{H}_{\mathcal{S}}$ commutes with $\mathcal{H}_{\mathcal{S}}$, $\mathcal{H}_{\mathcal{G}}=(\mathcal{H}_{\mathcal{G}}-\tau\mathcal{H}_{\mathcal{S}})^2$, $\mathcal{H}_{\mathcal{S}}=(\mathcal{H}_{\mathcal{S}})^2$, else go to (b).\\
\quad(b) let $\mathcal{H}_{\mathcal{G}}^{'}=\mathcal{H}_{\mathcal{G}}-\tau\mathcal{H}_{\mathcal{S}}$, $\mathcal{H}_{\mathcal{S}}^{'}=\mathcal{H}_{\mathcal{S}}$,
and update $f_n$ to new cost function $R_1^{'}$, where $\tau$ is a parameter.\\
\textit{step 5:} Perform Setp 3 for a searched parameter $\tau$.\\
\textbf{Output:} eigenstates $|\varphi_{0}\rangle,|\varphi_{1}\rangle,\cdots$ with eigenvalues $\lambda_{\min}=\lambda_0=f_0\leq \lambda_1\leq\cdots\leq\lambda_{\max}$.
\end{algorithm}

In step 1, given a series of parameter vectors $\vec\theta=(\theta_1,\cdots,\theta_L)$, the quantum circuit $U$ is defined as
\begin{equation}
\begin{aligned}
U(\vec\theta)&=U_L(\theta_L)U_{L-1}(\theta_{L-1})\cdots U_1(\theta_1)
\end{aligned}
\end{equation}
with $L$ components. Mathematically, after preparing an initial N-qubit state $|0\rangle^{\otimes N}$, the generated quantum state is defined as
\begin{equation}
\begin{aligned}
|\varphi\rangle=\Pi_{i=1}^LU_i(\theta_i)|0\rangle^{\otimes N}.
\end{aligned}
\end{equation}
Note that the number of parameters and $N$ are logarithmically proportional to the dimension of the generated state $|\varphi\rangle$ \cite{HardwareVQE2017,PQC20181,PQC20182}. These parameterized quantum circuits has been shown significant potential in generative adversarial learning \cite{GAN20181,GAN20182} and quantum circuit Born machines \cite{BornMachine2018}.

In step 3, we show how to obtain the generalized eigenstate and corresponding generalized eigenvalue. Our results rely on the fact that the Rayleigh quotient \cite{Parlett1998}
\begin{equation}
\begin{aligned}
\mathcal{R}(|\varphi\rangle;\mathcal{G},\mathcal{S})=\frac{\langle\varphi|\mathcal{G}|\varphi\rangle}{\langle\varphi|\mathcal{S}|\varphi\rangle}, \langle\varphi|\mathcal{S}|\varphi\rangle\neq0
\end{aligned}
\end{equation}
is stationary at $|\varphi\rangle\neq0$ if and only if $(\mathcal{G}-\lambda \mathcal{S})|\varphi\rangle=0$ for some scalar $\lambda$ where $\mathcal{S}$ is positive definite \cite{Parlett1998}. Let $\mathcal{H_\mathcal{G}}=\mathcal{G}$ and $\mathcal{H_\mathcal{S}}=\mathcal{S}$ which also have the decomposition like Eq. (17). The first iteration obtains the generalized eigenstate with the lowest generalized eigenvalue by minimizing the $\mathcal{R}(|\varphi\rangle;\mathcal{G},\mathcal{S})$.

To find generalized eigenstates, we firstly introduce two cost functions for tackling two different cases. When $\mathcal{G}$ commutes with $\mathcal{S}$, we update the Hamiltonian $\mathcal{H}_{\mathcal{G}}=(\mathcal{H}_{\mathcal{G}}-\tau\mathcal{H}_{\mathcal{S}})^2$, $\mathcal{H}_{\mathcal{S}}=(\mathcal{H}_{\mathcal{S}})^2$, where $\tau$ is a parameter close to the energy of the generalized eigenstates, which turns the generalized eigenvalues into the ground state energy of updated Hamiltonian ($\mathcal{H}_{\mathcal{G}},\mathcal{H}_{\mathcal{S}}$). The following derivation ensure this modification provides all generalized eigenvalues. We consider the equation:
\begin{equation}
\begin{aligned}
(\mathcal{H}_{\mathcal{G}}-\tau\mathcal{H}_{\mathcal{S}})^2|\varphi\rangle&=(\mathcal{H}_{\mathcal{G}}^2+\tau^2\mathcal{H}_{\mathcal{S}}^2
-2\tau\mathcal{H}_{\mathcal{G}}\mathcal{H}_{\mathcal{S}})|\varphi\rangle\\
&=(\lambda+\frac{\tau^2}{\lambda}-2\tau)\mathcal{H}_{\mathcal{G}}\mathcal{H}_{\mathcal{S}}|\varphi\rangle\\
&=(\lambda-\tau)^2\frac{1}{\lambda}\mathcal{H}_{\mathcal{G}}\mathcal{H}_{\mathcal{S}}|\varphi\rangle\\
&=(\lambda-\tau)^2\mathcal{H}_{\mathcal{S}}^2|\varphi\rangle.
\end{aligned}
\end{equation}
The second equality uses the assumption that $\mathcal{G}$ commutes with $\mathcal{S}$ ($\mathcal{H}_{\mathcal{G}}\mathcal{H}_{\mathcal{S}}=\mathcal{H}_{\mathcal{S}}\mathcal{H}_{\mathcal{G}}$).
Therefore, the Rayleigh quotient is
\begin{equation}
\begin{aligned}
R_1=\frac{\langle\varphi|(\mathcal{H}_{\mathcal{G}}-\tau\mathcal{H}_{\mathcal{S}})^2|\varphi\rangle}{\langle\varphi|\mathcal{H}_{\mathcal{S}}^2|\varphi\rangle}
=(\lambda-\tau)^2.
\end{aligned}
\end{equation}
Clearly, since Eq. (23) is quadratic function of the variable $\tau$, the ground generalized eigenstate of the updated Hamiltonian is found on the unique minimum point.

However, the above approach is useless when it is applied to the general situation such as $\mathcal{G}\mathcal{S}\neq\mathcal{S}\mathcal{G}$. Our second alternative approach now is presented. We update the Hamiltonian to the following form
$$\mathcal{H}_{\mathcal{G}}^{'}=\mathcal{H}_{\mathcal{G}}-\tau\mathcal{H}_{\mathcal{S}},\mathcal{H}_{\mathcal{S}}^{'}=\mathcal{H}_{\mathcal{S}}.$$
The presented Hamiltonian induces a new cost function
\begin{equation}
\begin{aligned}
R_1^{'}&=\Bigg(\frac{\langle\varphi|(\mathcal{H}_{\mathcal{G}}-\tau\mathcal{H}_{\mathcal{S}})|\varphi\rangle}
{\langle\varphi|\mathcal{H}_{\mathcal{S}}|\varphi\rangle}\Bigg)^2
=\frac{(\langle\varphi|(\mathcal{H}_{\mathcal{G}}-\tau\mathcal{H}_{\mathcal{S}})|\varphi\rangle)^2}
{(\langle\varphi|\mathcal{H}_{\mathcal{S}}|\varphi\rangle)^2}\\
&=(\lambda-\tau)^2,
\end{aligned}
\end{equation}
which is calculated by performing QEE for updated Hamiltonian.

We next estimate the energy gap $\Delta=\lambda_{\max}-\lambda_{\min}$ by finding the minimum and maximum of Eq. (21). After estimating the energy interval $[\lambda_{\min},\lambda_{\max}]$, one tunes the parameter $\tau$ from $\lambda_{\min}$ to $\lambda_{\max}$ with a step size $d$, for example, $d=\frac{\lambda_{\max}-\lambda_{\min}}{1000}$. For each parameter $\tau$, one can estimate the minimum of cost function $R_{1}$($R_1^{'}$) by measuring the corresponding expectation values. When the experimental minimal value of cost function $R_{1}$($R_1^{'}$) tends to zero, it indicates $\tau$ is close to the generalized eigenvalue of $(\mathcal{G},\mathcal{S})$. Notice that the minimum corresponds to a generalized eigenvalue and a optimal parameter vector $\vec\theta_{opt}$ which is applied for preparing the generalized eigenstates by requesting the variational circuits $U(\vec\theta_{opt})$. The searched method is similar to the idea of \cite{Wang1994,Shen2017}. Finally, we sort the generalized eigenvalues and output all eigenstates via the unitary circuit in step 1.

The main result of this subsection is the following theorem.

\textit{Theorem 3.} For a Hermitian matrix pencil $(\mathcal{G},\mathcal{S})$ with invertible matrix $\mathcal{S}$, let $\epsilon>0$ be a precision parameter. Algorithm 2 has the coherence time $O(1)$ that outputs all generalized eigenstates of the following generalized eigenvalue problem:
$$\mathcal{G}|\varphi\rangle=\lambda \mathcal{S}|\varphi\rangle,$$
requiring $\tilde{O}(1/\epsilon^2)$ repetitions, where $\mathcal{G},\mathcal{S}\in\mathcal{R}^{n\times n}$ and $|\varphi\rangle$ is the generalized eigenstate corresponding to the generalized eigenvalue $\lambda$.

\textit{Proof.} The time the quantum computer remain coherent is $O(1)$ which is determined by the extra depth of used circuit for preparing the parameterized state. If the desired error is at most $\epsilon$, the cost of the expectation estimation of local Hamiltonian $\mathcal{H}^m$ is $O(|\max\{h_{12\cdots}^{ij\cdots}\}|^2/\epsilon^2)$ repetitions of the preparation and measurement procedure \cite{VQE2014}. The overall generalized eigenstates can be prepared via $n$ times queries for the parameter quantum circuit and $M$ Hamiltonian items. Thus, we require $\tilde{O}(1/\epsilon^2)=O(nMP|\max\{h_{12\cdots}^{ij\cdots}\}|^2/\epsilon^2)$ samples from the parameterized circuit with coherence time $O(1)$, where the constant $P$ is determined by the classical minimization method used and the $\tilde{O}$ suppresses constant items. Due to the fact that our cost function occurs on the quantum computer, our Algorithm 2 has a speedup over classical cost evaluation. $\hfill\blacksquare$

With the assistance of Theorem 3, only replacing $\mathcal{G}$, $\mathcal{S}$ with $XQX^{\dag}$ and $XX^{\dag}$ can we find the $d$ lower eigenstates $\{|a_i\rangle\}_{i=0}^{d-1}$ as the column of the projection matrix $A$ with runtime $\tilde{O}(1/\epsilon^2)$. Note that $XQX^{\dag}$ and $XX^{\dag}$ are positive definite matrix in $\mathcal{R}^{D\times D}$ \cite{NPE2005}. One calculates these two Hermitian matrices by matrix multiplication algorithm on classical computer \cite{MM1990}. Assuming that these two matrices can be regarded as raw-computable Hamiltonian, Berry et. al \cite{Berry2007} have shown that $XQX^{\dag}$ and $XX^{\dag}$ may be decomposed as a sum of at most $O(6D^2)$ $1$-sparse matrices each of which is efficiently simulated in $O(\log_2 D)$ queries to the Hamiltonian.
\subsection{Extract the lower-dimensional manifold}
We now extract the low-dimension manifold based upon the projection matrix $A$.
The embedding state is given as:
\begin{equation}
\begin{aligned}
|x_i\rangle\mapsto|y_i&\rangle=A^{\dag}|x_i\rangle,\\
&A=(|a_{0}\rangle,|a_{1}\rangle,\cdots,|a_{d-1}\rangle),
\end{aligned}
\end{equation}
where $|y_i\rangle$ is a $d$-dimensional vector and $A$ is a $D\times d$ matrix. Our qNPE maps arbitrary high dimensional vector to a lower-dimensional vector. Thus if one is given a test vector $|x_{test}\rangle$, then the embedding vector is $|y_{test}\rangle=A^{\dag}|x_{test}\rangle$.

Here, we propose two optional methods for the extraction of the embedding states. One of them is based on QSVD. Like \cite{QSVD2018}, an extended matrix is considered as
\begin{equation}
\tilde{A}=\begin{pmatrix}0 & A^{\dag}\\A &0\end{pmatrix}.
\end{equation}
Assume that $\tilde{A}$ has eigenvalue decomposition
\begin{equation}
\tilde{A}=\sum_{j}\sigma_j|\tilde{u}_{+}^j\rangle\langle\tilde{u}_{+}^j|
-\sigma_j|\tilde{u}_{-}^j\rangle\langle\tilde{u}_{-}^j|
\end{equation}
with singular value decomposition $A^{\dag}=\sum_{j}\sigma_j|u_j\rangle\langle v_j|$, where
$|\tilde{u}_{\pm}^j\rangle=\frac{1}{\sqrt{2}}
(|0\rangle|u_j\rangle\pm|1\rangle|v_j\rangle)$. We then perform quantum phase estimation on the initial state $|0,x_i\rangle|0,\cdots,0\rangle$ and obtain a state
\begin{equation}
\sum_{j}\alpha_j^{\pm}|\tilde{u}_{\pm}^j\rangle|\pm\frac{\sigma_j}{d+D}\rangle|0\rangle,
\end{equation}
where $\alpha_j^{\pm}=\pm\frac{\langle v_j|x_i\rangle}{\sqrt{2}}$. Performing a Pauli operator $\sigma_z$ on the flag qubit and applying $R_y(2\arcsin\frac{\sigma_j}{d+D})$ on an ancilla qubit $|0\rangle$, we generate a state
\begin{equation}
\sum_{j}\alpha_j^{+}|0\rangle|u_j\rangle\left[\frac{\sigma_j}{d+D}|1\rangle+\sqrt{1-\Bigg(\frac{\sigma_j}{d+D}\Bigg)^2}|0\rangle\right].
\end{equation}
To this end, we project onto the $|u_j\rangle$ part and measure the final qubit to 1 resulting in a state
\begin{equation}
\sum_{j}\frac{\sigma_j}{d+D}\alpha_j^{+}|u_{j}\rangle\propto \sum_{j}\sigma_j|u_j\rangle\langle v_j|x_i\rangle=A^{\dag}|x_i\rangle.
\end{equation}
Repeating the above process $M$ times, the embedding state $|y_0\rangle,|y_1\rangle,\cdots,|y_{M-1}\rangle$ will be prepared with error $\epsilon$ in time $O(M\log_2^2(D+d)\|A\|_{\max}^{2}/\epsilon^3)$.

Alternatively, another approach is based on the well-known swap test \cite{swaptest}. Since the embedding low-dimensional data is
\begin{equation}
\begin{aligned}
|y_i\rangle=A^{\dag}|x_i\rangle&=(\langle a_{0}|x_i\rangle,\langle a_{1}|x_i\rangle,\cdots,\langle a_{d-1}|x_i\rangle)^{\dag},
\end{aligned}
\end{equation}
we convert formula (25) into a computation of inner product item $\langle a_{k}|x_i\rangle$. The swap test calculates the square of the inner product by the expectation of operators. But here the magnitude and sign of these inner products are also required. Fortunately, the inner product can be estimated with $O(\log_2 D)$ number of measurements \cite{Liu2018,zhao2019}. The embedding low-dimensional vector can be computed using resources scaling as $O(Md\log_2 D)$. In summary, this two approaches help one to obtain the embedding vectors with quantum (classical) form which can be directly applied in other quantum (classical) machine learning process.
\begin{figure*}[ht]
\subfigure[]{\includegraphics[scale=0.7]{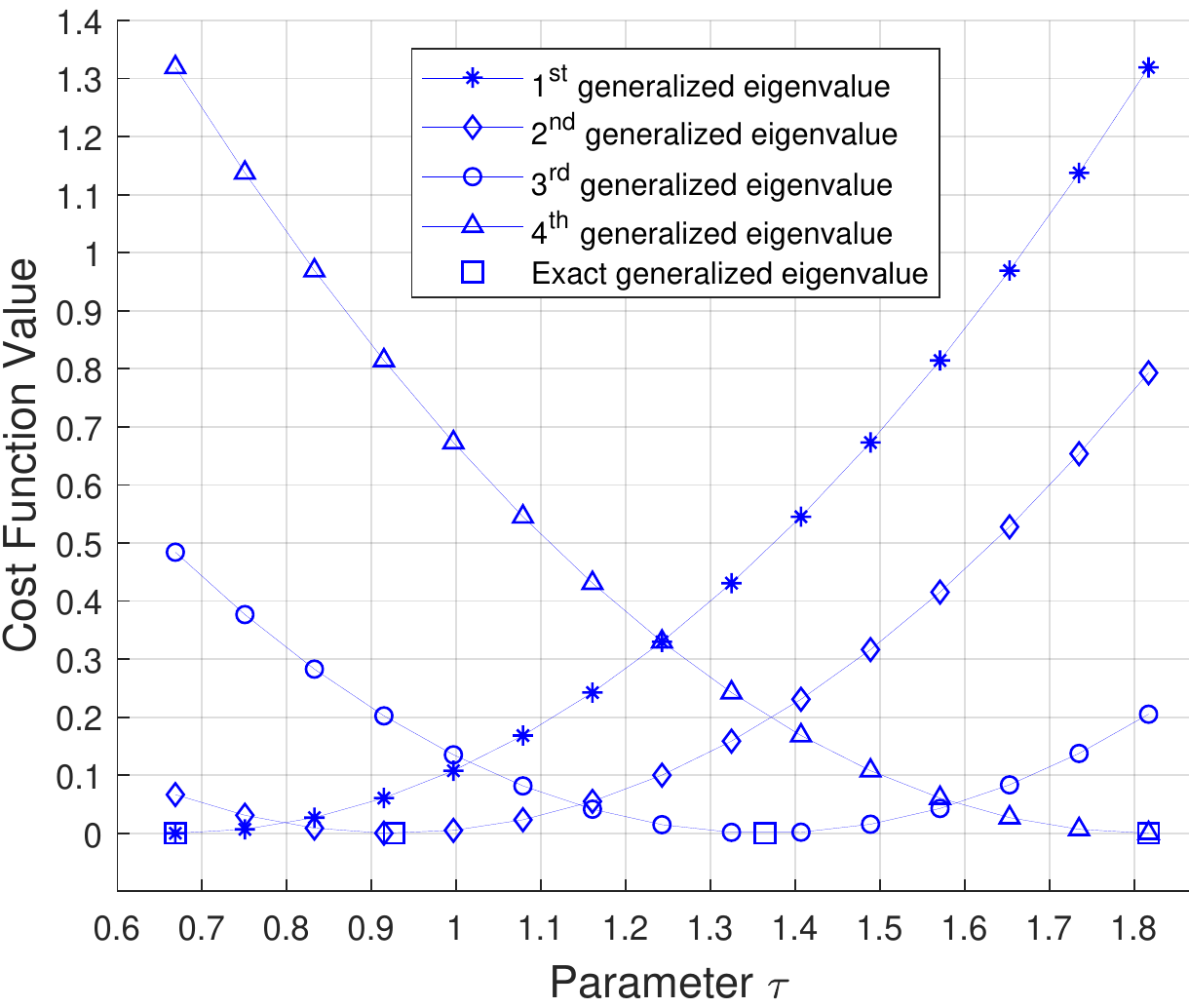}}
\subfigure[]{\includegraphics[scale=0.7]{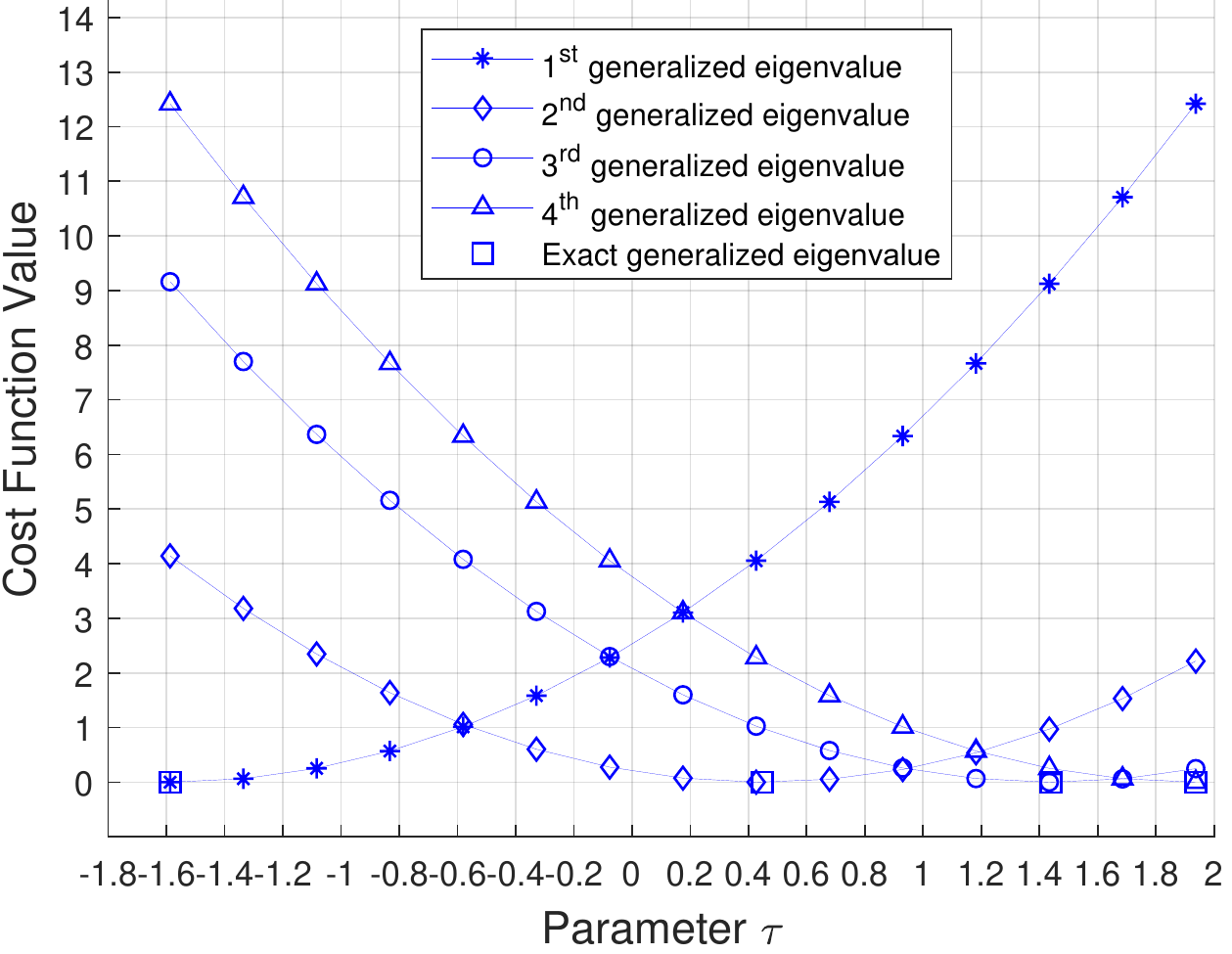}}
\caption{Search process for generalized eigenvalues of (a) Example 1 and (b) Example 2. After estimating the energy interval $[\lambda_{\min},\lambda_{\max}]$, one tunes the parameter $\tau$ from $\lambda_{\min}$ to $\lambda_{\max}$ with a step size $d$, for example, $d=\frac{\lambda_{\max}-\lambda_{\min}}{1000}$. For each parameter $\tau$, one can estimate other two generalized eigenvalues by the minimizing the cost function $R_{1}$($R_{1}^{'}$). The minimum of each cost function corresponds to a $\tau$ which is equal to a generalized eigenvalue with optimal vector $\vec\theta_{opt}$.}
\end{figure*}
\subsection{Numerical simulations and performance analysis}
In this subsection, we conduct a numerical experiment to simulate the proposed VQGE. For the implementation, we consider the following two $32\times32$ matrices (using $5$-qubits).

\textit{Example 1:}
\begin{equation}
\begin{aligned}
&\mathcal{G}_1=\mathds{1}+0.2\sigma_1^1\otimes\sigma_3^2+0.5\sigma_1^1\otimes\mathds{1},\\
&\mathcal{S}_1=\mathds{1}+0.441\sigma_1^1\otimes\sigma_3^2+0.3939\sigma_1^1\otimes\mathds{1},
\end{aligned}
\end{equation}
which has four different generalized eigenvalues $\lambda_1=0.6685,\lambda_2=0.9265,\lambda_3=1.3643,\lambda_4=1.8171$. In example 1, we only consider the case when $\mathcal{G}$ commutes with $\mathcal{S}$.

\textit{Example 2:}
\begin{equation}
\begin{aligned}
&\mathcal{G}_2=\mathds{1}+0.63\sigma_1^1\otimes\sigma_3^2\otimes\mathds{1}
+1.2\sigma_1^1\otimes\mathds{1}+0.2\sigma_3^1\otimes\mathds{1},\\
&\mathcal{S}_2=\mathds{1}+0.1741\sigma_1^1\otimes\sigma_3^2\otimes\mathds{1}
+0.2981\sigma_1^1\otimes\mathds{1}.
\end{aligned}
\end{equation}
Example 2 gives a general case for $\mathcal{G}_2\mathcal{S}_2\neq\mathcal{S}_2\mathcal{G}_2$ which also has four different generalized eigenvalues $\lambda_1=-1.5872,\lambda_2=0.4480,\lambda_3=1.4396,\lambda_4=1.9370$.

Here, we utilize a common variational circuit $U(\vec\theta)$ introduced in \cite{HardwareVQE2017,Supervised2019}. The variational circuit $U(\vec\theta)$ is parametrized by $\vec\theta\in\mathcal{R}^{2n(L+1)}$, $U(\vec\theta)=U_{R}^{L}(\theta_L)U_{ent}\cdots U_{R}^{2}(\theta_2)U_{ent}U_{R}^{1}(\theta_1)$ which contains $L$ layers. We alternate layers of entangled gates $U_{ent}=\Pi_{(i,j)}Z(i,j)$ with full layers of single-qubit rotations $U_{R}^t(\theta_t)=\otimes_{i=1}^nU(\theta_i^t)$ with $U(\theta_i^t)\in\textrm{SU}(2)$. The entangled unitary $U_{ent}$ consists of the controlled $Z$ gates applied on the $i$ and $j$ qubits. This short-depth circuit can generate any unitary if sufficiently many layers $L$ are applied \cite{Supervised2019}. Appendix \ref{appendixA} presents a detailed analysis on this ansatz and our experiments setting.

The experiment's results of the VQGE implementation are shown in Fig. 1. The first and the last eigenvalue is estimated by finding the minimum and maximum of Eq. (21). Other generalized eigenvalues is found by scan $\tau$ from $\lambda_{\min}$ to $\lambda_{\max}$ with a step size $d$, for example, $d=\frac{\lambda_{\max}-\lambda_{\min}}{1000}$. For each parameter $\tau$, one can estimate the minimum of cost function $R_{1}$($R_{1}^{'}$) by measuring the corresponding expectation values. The minimum of each cost function corresponds to a $\tau$ which is equal to a generalized eigenvalue. The cost function $R_{1}$ is applied for example 1 and $R_{1}^{'}$ for example 2 according to the analysis in Sec. II C. Finally, once all optimal parameters are determined, we obtain the generalized eigenvalues via the expectation values of different Hamiltonian.
\begin{figure}[htbb]
\centering
\includegraphics[scale=0.3]{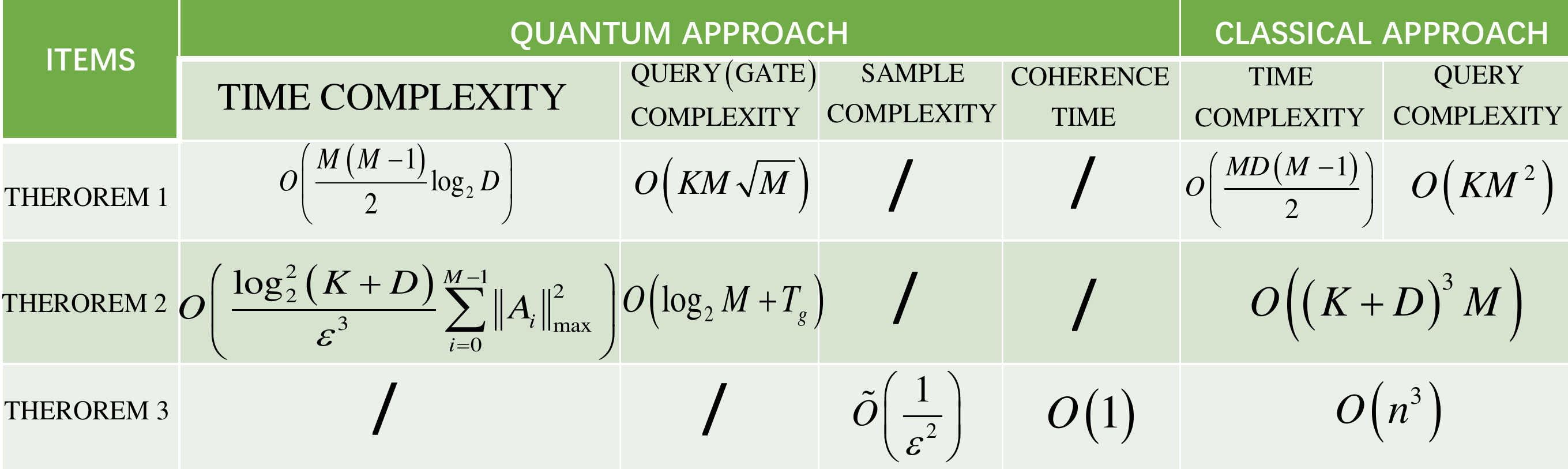}
\caption{Required resource complexity of quantum and classical methods.}
\end{figure}

Fig. 2 shows the required resources of quantum and classical methods. Classically, performing Theorem 2 have a runtime $O((K+D)^3M)$. The runtime complexity of solving the generalized eigenvalue problem is of order $O(n^3)$ on classical computation devices \cite{Golub2012}. And we have explained the required classical resources of Theorem 1 and 3 in subsection A and C.

However, as Aaronson has pointed out in \cite{Aaronson2015read}, it is also not clear whether the qNPE achieves an exponential speedup over classical part for practically instances of a dimensionality reduction problem. It is due to the fact that one requires a more efficient state preparation technique for uploading the classical points into a quantum states before the swap test and the QSVD. One recent result indicates that there has a quantum-inspired classical recommendation system exponentially faster than previous classical systems \cite{Tang2019a}. In this work, the qNPE is based on the amplitude encoding using $\log_2 D$ qubits for a $D$-dimensional classical data point. Currently, the preparation of arbitrary quantum states is still a nontrivial topic, although some techniques have been developed such as the well-conditioned oracle \cite{Clader2013preconditioned} and the quantum RAM \cite{qRAM2008}. Thus, we need to treat the exponential speeds carefully for machine learning problem.
\section{Quantum Local Discriminant Embedding}
In this section, based on the variational quantum generalized eigenvalues (VQGE), we develop a quantum algorithm for pattern classification which preserves the local manifold. This algorithm is a quantum version of local discriminant embedding \cite{LDE2005} (qLDE). The task is to classify a high-dimensional vector into one class, given $M$ data points of the form $\{(x_i,\mathbf{c}_i):x_i\in\mathcal{R}^{D},\mathbf{c}_i\in\{1,2,\cdots,P\}\}_{i=0}^{M-1}$ where $\mathbf{c}_i$ depends on the class to which $x_i$ belongs. Fig. 3 shows the expected effect of local discriminant embedding. After finding an associated submanifold of each class, the qLDE separates the embedded data points into a multi-class lower-dimensional Euclidean space.

First of all, one needs to construct two neighborhood graphs: the intrinsic graph $G_{w}$ (within-class graph) and the penalty graph $G_{b}$ (between-class graph). For each data point $x_i$, we define a subset $\mathcal{N}_{w,i,K}$ ($\mathcal{N}_{b,i,K^{'}}$) which contains the $K$ ($K^{'}$) neighbors having the same (different) class label with $x_i$. For graph $G_{w}$, we consider each pair of $x_i$ and $x_j$ with $\mathbf{c}_i=\mathbf{c}_j$. An edge is added between $x_i$ and $x_j$ if $x_j\in\mathcal{N}_{w,i,K}$. To construct $G_{b}$, likewise, we consider each pair of $x_i$ and $x_j$ with $\mathbf{c}_i\neq \mathbf{c}_j$. An edge is added if $x_j\in\mathcal{N}_{b,i,K^{'}}$. Theorem 1 can help us to finish the construction of $G_{w}$ and $G_{b}$ by finding $K$ ($K^{'}$) neighbors.
\begin{figure}[h]
\centering
\includegraphics[width=2.5in]{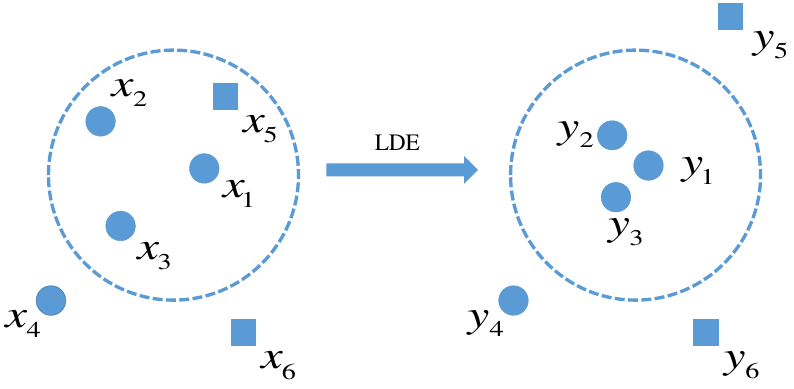}
\caption{The expected effect of LDE \cite{Dornaika2013}. The point $x_1$ have three neighbors $(x_2,x_3,x_5)$. The points with same shape belong to the same class. The within-class graph connects nearby points with the same label. The between-class graph connects nearby points with different labels. After LDE, the local margins between different classes are maximized, and the distances between local homogeneous samples are minimized.}
\end{figure}

Next, we determine the weight matrix $W_{w(b)}=(W_{w(b),ij})$ of graph $G_w$($G_b$) by the following convex optimization formulation:
\begin{equation}
\begin{aligned}
&\min\quad \sum_{i=0}^{M-1}\|x_i-\sum_jW_{w(b),ij}x_j\|^2\\
&\textrm{such}\quad\textrm{that}\quad \sum_{j=0}^{{K-1}}W_{w(b),ij}=1,i=0,1,\cdots,M-1.
\end{aligned}
\end{equation}
Theorem 2 prepares two weight states
$$|\psi_{W_w}\rangle=\sum_{i=0}^{M-1}|\omega_{wi}\rangle|i\rangle,\quad|\psi_{W_b}\rangle=\sum_{i=0}^{M-1}|\omega_{bi}\rangle|i\rangle,$$
with error at most $\epsilon$ in runtime $$O\Bigg(\frac{\log_2^2(K(K^{'})+D)}{\epsilon^3}\sum_{i=0}^{M-1}\|A_i\|_{max}^2\Bigg).$$
The required gate resource count is $O(T_g+\log_2 M))$.

We next turn to find the projection matrix $A$ that maximizes the local margins among different classes and pushes the homogenous samples closer to each other \cite{Dornaika2013}. The overall process corresponds to the below mathematical formula:
\begin{equation}
\begin{aligned}
&\min_{A}\frac{1}{2}\quad \sum_{ij}\|A^\dag(x_i-x_j)\|^2W_{w(b),ij}.
\end{aligned}
\end{equation}
After simple matrix algebra (seeing details in \cite{LDE2005}), the columns of the projection matrix $A$ are the generalized eigenvectors with the $d$ largest different eigenvalues in
\begin{equation}
\begin{aligned}
T_b|a\rangle=\lambda T_w|a\rangle.
\end{aligned}
\end{equation}
where $T_{w(b)}=X(I_{w(b)}-W_{w(b)})X^{\dag}$, $X=(x_0,x_1,\cdots,x_{M-1})$ and $I_{w(b)}$ is a diagonal matrix with $I_{w(b),ii}=\sum_{j}W_{w(b),ij}$. Then, we apply Theorem 3 to obtain the $d$ generalized eigenvectors with $d$ largest different eigenvalues of (36).

Once we have learned the projection matrix $A$ using qLDE, the embedding state is obtained via the following transformation:
$$
\begin{aligned}
|x_i\rangle\mapsto |&y_i\rangle=A^{\dag}|x_i\rangle,\\
&A=(|a_{0}\rangle,|a_{1}\rangle,\cdots,|a_{d-1}\rangle),
\end{aligned}
$$
where $|y_i\rangle$ is a $d$-dimensional vector and $A$ is a $D\times d$ matrix. Similarly, a given test state is projected to a state $|y_{test}\rangle=A^{\dag}|x_{test}\rangle$. Finally, quantum nearest neighbor algorithm \cite{Wiebe2015} is directly applied on multi-class classification tasks by computing the distance metrics between the test point $|y_{test}\rangle$ and other training points with a known class label. For example, for a given two clusters $\{U\}$ and $\{V\}$, if
\begin{equation}
\begin{aligned}
\min_{u\in \{U\}}\quad D(|y_{test}\rangle,|u\rangle)\leq\min_{v\in \{V\}}\quad D(|y_{test}\rangle,|v\rangle),
\end{aligned}
\end{equation}
then we can assign $|y_{test}\rangle$ to cluster class $\{U\}$, where $D$ denotes the trace distance. The classification performance show exponential reductions with classical methods \cite{Wiebe2015}.

\section{Conclusions and discussion}
In conclusion, this work presented qNPE and qLDE for dimensionality reduction and classification. Both of them preserve the local structure of the manifold space in the process of dimensionality reduction. We demonstrated that qNPE achieves an exponential advantage over the classical case since every steps of qNPE have an exponential speedup. The performance of qLDE on classification tasks is also competitive with classical analog.

Along the way, we developed two useful subroutines in machine learning and scientific computation. The first one is quantum $K$ nearest neighborhood search which finds $K$ lowest values in an unordered set with $O(K\sqrt{N})$ times. It may help us sort an unordered list with an upper bound $O(N\sqrt{N})$. Another subroutine is a variation hybrid quantum-classical algorithm for solving the generalized eigenvalue problem. In electronic structure calculations, for instance, the electron density can be computed by obtaining the eigenpairs $(E_m,\Psi_m)$ of the Schr\"{o}dinger-type eigenvalue problem $\mathcal{H}\Psi_m=E_m\mathcal{S}\Psi_m$ with different discrete energies $E_m$, where $\mathcal{H}$ denotes the Hamiltonian matrix and $\mathcal{S}$ is a symmetric positive matrix \cite{Polizzi2009}. Our variational quantum generalized eigenvalue solver can obtain the eigenpairs $(E_m,\Psi_m)$ in runtime $\tilde{O}(1/\epsilon^2)$ with error $\epsilon$ independent of the size of the Hamiltonian. Notice that our VQGE does not use the Hamiltonian simulation, amplitude amplification and phase estimation. We have performed numerical experiments solving the generalized eigenvalue problems with size $2^5\times2^5$. In the main text, we consider the noiseless evolution of quantum states. Actually noise resilience may be a general phenomenon when one applies variational quantum algorithms (including our VQGE) on NISQ computer \cite{mcclean2016the,shatri2019noise}. Although we have considered measurement noise and given an error bound in Appendix \ref{appendixB}, this problem is still required to be considered in our near future work.

While we have presented two algorithms for dimensionality reduction and classification, some questions still need further study. For example, it is a big challenge that how to construct the Hamiltonian $X^{\dag}QX(X^{\dag}X)$ from the entanglement state $|\psi_W\rangle=\sum_{i=0}^{M-1}|\omega_i\rangle|i\rangle.$ Finally, as the effect of artificial neural networks to the quantum many-body problem \cite{Carleo2017}, it would be interesting to investigate if our algorithms can also reduce the exponential complexity of the many-body wave function down to a tractable computational form.

\textit{Acknowledgments} The authors thank anonymous referees and editor for useful feedback on the manuscript. This work is supported by NSFC (11775306) and the Fundamental Research Funds for the Central Universities (18CX02035A, 18CX02023A, 19CX02050A).
\begin{appendix}
\section{Variational Ansatz}\label{appendixA}
In this section, we analyze two different variational ansatz circuits which are performed to generate the trial state.

The first variational circuit is the product ansatz. For example, an $n-$qubits quantum state $|\varphi(\vec\theta)\rangle$ is represented by a tensor product
\begin{equation}
\begin{aligned}
|\varphi(\vec\theta)\rangle&=U(\vec\theta)|0\rangle^{\otimes n}
=\otimes_{i=0}^{n-1}R_y(\theta_i)|0\rangle^{\otimes n}\\
&=\bigotimes_{j=0}^{n-1}\Bigg(\cos\frac{\theta_j}{2}|0\rangle+\sin\frac{\theta_j}{2}|1\rangle\Bigg).
\end{aligned}
\end{equation}
The vector $\vec\theta$ is defined as $\vec\theta=(\theta_0,\theta_1,\cdots,\theta_{n-1})^{\dag}$ and the rotation operator is $R_y=e^{-i\theta Y/2}$. Since the trial state $|\varphi(\vec\theta)\rangle$ is a separated state, it can only be applied on special matrices which have separated eigenvectors. However, in general, the eigenvectors of a given matrix may be an entangled eigenvector. Thus, one can choose the following variational ansatz circuit $U(\vec\theta)$ to prepare an entangled trial state $|\varphi(\vec\theta)\rangle=U(\vec\theta)|0\rangle^{\otimes n}$.

The variational circuit $U(\vec\theta)$ is represented as
\begin{equation}
U(\vec\theta)=U_{R}^{L}(\theta_L)U_{ent}\cdots U_{R}^{2}(\theta_2)U_{ent}U_{R}^{1}(\theta_1),
\end{equation}
where $U_{ent}=\Pi_{(i,j)}Z(i,j)$, $U_{R}^t(\theta_t)=\otimes_{i=1}^nU(\theta_i^t)$ with $U(\theta_i^t)\in\textrm{SU}(2)$. The entangled unitary $U_{ent}$ consists of the controlled $Z$ gates. In \cite{HardwareVQE2017}, Havl\'i\v{c}ek \textit{et al}. have used the variational quantum circuit to solve a classification problem of supervised machine learning.

In our experiment, the trial state $|\varphi(\vec\theta)\rangle$ is prepared by repeating one time after applying a rotation operator $R_y=e^{-i\theta Y/2}$ on five qubits. The variational circuit is shown in Fig. 4.
\begin{figure}[hb]
\centering
\includegraphics[scale=1.5]{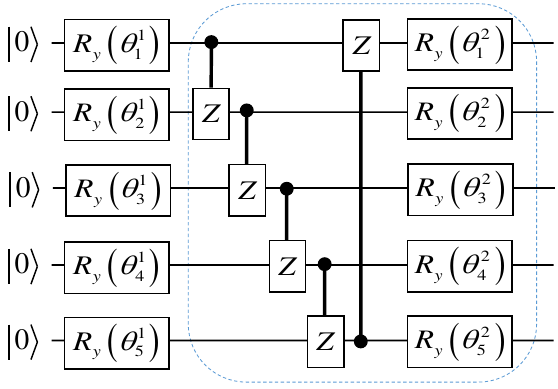}
\caption{The variational circuit for preparing the trial state $|\varphi(\vec\theta)\rangle$, where $\vec\theta=(\theta_1^1,\cdots,\theta_5^1,\theta_1^2,\cdots,\theta_5^2)^{\dag}$. The dashed box indicates the repeated block.}
\end{figure}
\section{Error Analysis}\label{appendixB}
In the main text, we consider the ideal situation without any noise. However, noise may be a general phenomenon when one applies variational quantum algorithms (including our VQGE) on NISQ computer. For example, such as measurement noise, gate noise, and Pauli channel noise. In this section, we only consider the measurement noise and find an error bound of generalized eigenvalues.

Let $\delta>0$ denotes the error in estimating expectation value $\langle\mathcal{G}\rangle$, $\langle\mathcal{S}\rangle$. We easily obtain the error $\mathcal{E}$ of the generalized eigenvalue of matrix pair $(\mathcal{G},\mathcal{S})$ is
\begin{equation}
\begin{aligned}
\mathcal{E}&=\left|\frac{\langle\mathcal{G}\rangle+\delta}{\langle\mathcal{S}\rangle+\delta}-\frac{\langle\mathcal{G}\rangle}{\langle\mathcal{S}\rangle}\right|
\leq\left|\frac{\langle\mathcal{G}\rangle+\delta}{\langle\mathcal{S}\rangle}-\frac{\langle\mathcal{G}\rangle}{\langle\mathcal{S}\rangle}\right|\\
&\leq\frac{\delta}{\langle\mathcal{S}\rangle}\leq\frac{\delta}{\lambda_{\min}^{\mathcal{S}}},
\end{aligned}
\end{equation}
where $\lambda_{\min}^{\mathcal{S}}$ is the minimum eigenvalue of $\mathcal{S}$. Thus, the error $\mathcal{E}$ has an upper bound $O(\delta/\lambda_{\min}^{\mathcal{S}})$. If we require this error to be of $O(\zeta)$, we need to take the measurement error to be $\delta=O(\lambda_{\min}^{\mathcal{S}}\zeta)$.
\end{appendix}


\begin{thebibliography}{99}
\bibitem{nearest} The set $\mathcal{N}_i$ contains $K$ vector states which also reside in the $M$ quantum states set.
\end{thebibliography}
\end{document}